\documentclass[reqno,12pt]{amsart}
\usepackage{amsmath,amssymb}
\usepackage[utf8]{inputenc}
\usepackage[left=2.6cm,top=2.6cm,right=2.6cm,bottom=2.6cm]{geometry}
\usepackage{epsfig}

\usepackage{graphicx}
\usepackage{subfigure}

\usepackage{epsfig}
\usepackage{bm}
\usepackage{graphicx}

\usepackage{makecell}
\usepackage[font=small,margin=1em]{caption}
\usepackage{tikz}
\usepackage{subfigure}

\usepackage{enumitem}
\usepackage{booktabs}

\newcommand{\sm}{\setminus}

\newcommand{\calG}{{\mathcal G}}

\newcommand{\calC}{\mathcal{C}}
\newcommand{\calD}{\mathcal{D}}

\renewcommand{\subset}{\subseteq}

\usepackage{array}

\title{Clique Counts for Network Similarity}

\author[A.\ Bonato]{Anthony Bonato}
\author[Z.\ Zhang ]{Zhiyuan Zhang}
\address[A1,A2]{Toronto Metropolitan University, Toronto, Canada}

\email[A1]{(A1) abonato@torontomu.ca}
\email[A2]{(A2) zhiyuan.zhang@torontomu.ca}

\begin{document}

\begin{abstract}

Counts of small subgraphs, or graphlet counts, are widely applicable to measure graph similarity. Computing graphlet counts can be computationally expensive and may pose obstacles in network analysis. We study the role of cliques in graphlet counts as a method for graph similarity in social networks. Higher-order clustering coefficients and the Pivoter algorithm for exact clique counts are employed. 
\end{abstract}

\maketitle

\section{Introduction}

Graph similarity is a central topic in the ever-expanding, interdisciplinary field of complex networks. Quantifying similarity between networks is crucial for revealing their latent structures and in model selection. On the applied side, graph similarity is widely used in many areas, such as in recommender systems in social network analysis, accelerating drug discovery, and understanding the structural similarity of biological molecules. 

One approach to graph similarity is to use counts of small subgraphs as a measure of similarity. Small subgraphs are also called \emph{graphlets} or \emph{motifs}. Graphlets have found wide application in many fields, such as the biological sciences, social network analysis, and character networks, often coupled with machine learning paradigms; see, for example, \cite{shford2022online,b16,borgwardt2020graph,feng2022motif,hurshman2012model,milo2002network,prvzulj2007biological,shervashidze2009efficient,sinha2022impact,yanardag2015deep,zhao2023intrinsic}. A challenge with graphlet counts is that they are often computationally expensive to compute exactly, especially for large networks.

We consider graph similarity via counts of cliques in networks. The present work also emphasizes the roles of cliques in complex networks and uses their counts as a measure of similarity. Cliques are simplified representations of highly interconnected structures in networks. For example, a clique in the Instagram social network consists of accounts linked via friendship or mutual interests. A recent model considered an evolving network model defined by growing cliques; the model simulated many properties found in social networks, such as densification power laws and the small world property; see \cite{bonato2022frustum}. Another recent model \cite{fox2020finding} proposed a new distribution-free model for social networks based on cliques. Cliques have also been studied from the point of view of their densification in evolving networks; see \cite{Pi2023cliquedensification}.

While cliques are pervasive in networks, we expect to find fewer of them in sparse networks. As such, our work is only broadly applicable to real-world networks with many cliques. In studies such as \cite{leskovec2005graphs}, social networks were found to densify and have rich community structure. One consequence is that social networks have dense subgraphs (cliques or cliques missing a small number of edges) corresponding to small communities. 

Our main goal in the present paper is to show that clique counts perform as reliably as other, more sophisticated graph similarity measures in certain networks, such as social networks. Recent work by Jain and Seshadhri \cite{jain2020power} proposed the Pivoter algorithm for exact clique count, which is well-suited for our empirical study. Studies such as \cite{Pi2023cliquedensification} have continued this work. As referenced earlier, an advantage of using clique counts is that they are computationally inexpensive. 

The paper is organized as follows. We define clique profiles in Section~2, providing definitions and relevant notation. We also consider a higher-order version of the clustering coefficient, which is another feature of our graph similarity approach. Section~3 focuses on our methods using clique profiles as measures of graph similarity. We analyze data sets from various domains and show that clique profiles perform as well as existing graph similarity methods in several cases. We conclude with a summary of the work and various open problems.

All graphs we consider are undirected. The clique (or complete graph) of order $n$ is denoted $K_n.$ If $S$ is a set of nodes in a graph $G$, then we write $G[S]$ for the subgraph induced by $S.$ For a node $v$, the {\em neighborhood of $v$} is $N(v)$ and $\mathrm{deg}(v) = |N(v)|$ is the {\em degree of $v$}. For further background on complex networks, see \cite{bonato2008course}; for more background on graph theory, see \cite{west}.

\section{Clique Profiles}

The clique profile of a graph (defined precisely below) is a normalized vector that consists of clique counts of various orders. We will use graph datasets with labels, compute their clique profiles, and examine the classification ability of clique profiles to separate labels.  We also observe the change of the $k$-clustering coefficient, first defined in \cite{yin2018higher}, on various growing networks. 

We start by defining notation. Let $\calG_k$ be the set of all non-isomorphic graphs of order $k$, where the graphs are arbitrarily indexed. The {\em count} of $G_i\in\calG_k$ is the number of subsets $S\subset V$ so that $G[S]$ is isomorphic to $G_i$. The {\em graph $k$-profile} of a graph $G$ is the set of the relative frequency among the counts of isomorphism-types of graphs of order $k$ that are subgraphs of $G$; it may be viewed as the embedding of $G$ into vector space, where the $i$-th coordinate is $g_i/(\sum_{G_j\in\calG} g_j)$. For example, the space is 4-dimensional if $k=3$ and 11-dimensional if $k=4$.

Graph profiles have appeared in many works; see, for example, \cite{hurshman2012model,shervashidze2009efficient,yanardag2015deep}. 
In \cite{b16}, Bonato et al.\ applied the graph profiles to select which random graph model best fits character networks from novels. In \cite{bonato2014dimensionality}, graph profiles are applied to determine the dimensionality of networks from certain random graph models. Computing the counts of all graphlets remains expensive, and only inexact counts of a small percentage of a graph are feasible; see \cite{ribeiro2021survey}.

For integers $3\leq j\leq k$, let $C_j(G)$ denote the number of $j$-cliques in $G$ and let $\bm C_k(G)$ be the corresponding $(k-2)$-dimensional vector, where the $(j-2)$-th component is $C_{j}(G)$. The {\em $k$-clique profile of $G$}, denoted $\mathcal{C}_k(G)$, is the normalized vector with each $j$-th component being
$$\frac{C_j(G)}{||\bm{C}_k(G)||},$$ where $||\cdot||$ denotes the 2-norm. 
For triangle-free graphs, we define its $k$-clique profile as the zero vector. 

The {\em clustering coefficient} is a fundamental measurement in network analysis, defined in \cite{watts1998collective}; see \cite{bonato2008course} for further background. Fix $v\in V$, and let $E_v$ be the edge set of the induced subgraph $G[N(v)]$. The {\em local clustering coefficient} of $v$ is defined as
$$ c(v) =c_G(v)= \frac{2|E_v|}{\mathrm{deg}(v)(\mathrm{deg}(v)-1)} = \frac{|E_v|}{\binom{\mathrm{deg}(v)}2}.$$
The {\em average clustering} of a graph $G$ is 
$$\mathrm{acc}(G) = \frac 1n \sum_{v\in V} c(v). $$
The {\em global clustering coefficient} of a graph $G$ is 
$$ \mathrm{cc}(G) = \frac{\text{number of triangles in $G$}}{\text{number of paths of length 2 in $G$}} = \frac{\sum_{v\in V}|E_v|}{\sum_{v\in V} \binom{\mathrm{deg}(v)}2}.$$
The clustering coefficient of a node may be viewed as the probability of two neighbors of a node such that they are adjacent, and the clustering coefficient of a graph is the ratio between the number of triangles and the number of all triplets.
An alternative way to describe $c_G(v)$ is the proportion of 2-cliques (or edges) that involve $v$ that form a 3-clique. 

\begin{figure}
    \centering
\includegraphics[scale=0.5]{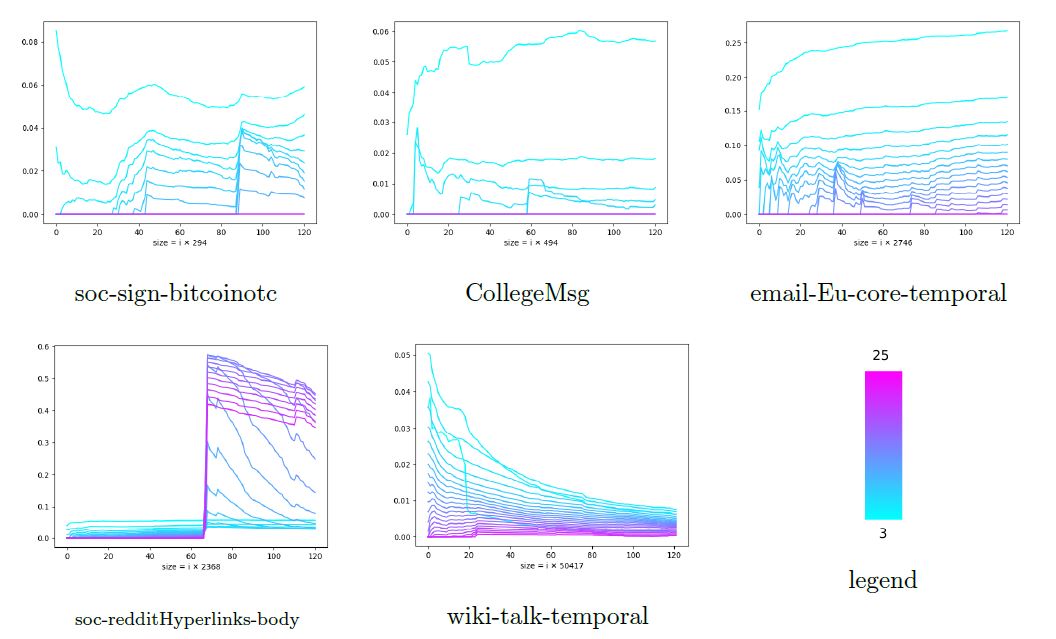}
    \caption{The change of the $k$-clustering coefficients of five networks retrieved from SNAP \cite{snap}, where each edge is associated with a time-stamp. 
We sort the edges and treat each large network as a sequence of 120 evolving networks with equal-size edge increments; every network has a fixed amount of edge growth from the previous network. We record their $k$-clustering coefficients up to $k=25$. 
The vertical axis indicates the $k$-clustering coefficient, the horizontal axis corresponds to the $i$-th network in the sequence, and the label indicates the size increment.    }

\end{figure}

Yin et al.\ \cite{yin2018higher} generalized this idea of clustering coefficients to cliques, which we describe next; see also \cite{lawford2020cliques}. Let $C_k(G; v)$ denote the number of $k$-cliques that involve node $v$. 
Consider the case where $G = K_n$, the complete graph of order $n>k$, and fix a node $v\in V(K_n)$. 
Note that the subgraph $K_n[V(K)\cup\{v'\}]$ forms a $k$-clique, for every $(k-1)$-cliques $K$ of $G$ that includes $v$ and some $v'\in N(v)\sm V(K)$. There are $k-1$ ways to form $K$; we then have that 
$$C_k(K_n; v) = C_{k-1}(K_n; v) (\mathrm{deg}(v) - ((k-1)-1)) / (k-1).$$
We also can observe that for a graph $G$ that 
$$ C_k(G; v) \leq \frac{C_{k-1}(G; v) (\mathrm{deg}(v) - k+2)}{k-1}.$$
Hence, we have the ratio 
$$ \mu_k(G; v) = \frac{C_k(G; v)(k-1)}{C_{k-1}(G; v) (\mathrm{deg}(v) - k+2)}\leq 1.$$ 
If $C_{k-1}(G;v) = 0$, then we simply define $\mu_k(G; v) = 0$. 
For convenience, we refer to $\mu_k(G; v)$ as the $k$-\emph{clustering coefficient} of the node $v$. 
Analogously, we can define the {\em $k$-clustering coefficient of $G$} as 
$$\mu_k(G) = \frac{(k-1)\sum_{v\in V}C_k(G; v)}{\sum_{v\in V} C_{k-1}(G; v) (\mathrm{deg}(v) - k + 2)}.$$
We observe that $\mu_3(G; v) = c_G(v)$, $\mu_3(G) = \mathrm{cc}(G)$, and $\mu_2(G; v) = 1$.

In \cite{yin2018higher}, they analyzed $\mu_k(G)$ for two network models: the Erd\H os-R\'enyi binomial random graph and the small-world graphs in \cite{watts1998collective}. They reported statistics such as the joint distributions of $(\mu_2(G), \mu_3(G))$ and the ratio of the nodes that are involved in 2-, 3-, and 4-cliques for real-world networks and graphs generated by complex network models. We consider the change of $k$-clustering coefficients in several large networks. See Figure~\ref{tab:my_label}. Surprisingly, we observed different patterns. 
For example, there is no clear sign such as all $k$-clustering coefficients would converge to zero or any other ratio; 
it is also not necessarily that $\mu_k(G)$ is lower bounded by $\mu_{k+1}(G)$, or vice versa, such as in the soc-redditHyperlinks-body dataset. 

\section{Experimental Design and Methods}

Our experiments aim to show that clique profiles are useful for graph similarity in certain networks. Our results will not surpass state-of-the-art results but closely match various benchmarks in the literature. Note that parameters used in clique profiles are obtainable by computing node-wise clique counts; we use Pivoter \cite{jain2020power} for this task.
We compute clique counts up to order ten and use the $k$-clique profile to embed each graph for $4\le k \le 10$. As a separate case, we also concatenate the global clustering coefficients to these clique profiles as inputs. 

We investigate the clique profiles on the following datasets, obtained from \cite{CON} and \cite{Morris2020tudataset}.
The networks in the dataset in \cite{Morris2020tudataset} are known as {\em ego-networks}; that is, networks of reasonable order sampled from a larger network. 
Sampling such networks usually follows the following two steps. 
First, sample a node in a network with some criteria (such as a label); it is usually more meaningful to sample one with a high degree number. Second, find the neighbor set of the sampled node that meets the criteria. We then have that every network is an induced subgraph comprising the sampled node and its neighbors. 
We briefly summarize statistics for the datasets in Table~\ref{tab:my_label}.
\begin{table}[h]
    \centering
    \begin{tabular}{|c|c|c|c|}\hline
        Dataset Name &  Labels & \makecell{\# of \\Labels} & \makecell{\# of Networks\\(per label)}  \\\hline\hline
        COLLAB & Subjects & 3 &2600/775/1625\\  
        IMDB-BINARY & Movie Genres & 2  & 500/500\\  
        IMDB-MULTI & Movie Genres & 3  &500/500/500\\  
        Github Stargazers & {Developer Communities} & 2 &5917/6808\\  
        Deezer Ego Nets& User Genders & 2  & 5470/4159\\  
        Survivor and Big Brother& TV Series & 2  & 21/37\\  \hline
    \end{tabular}
    \caption{Statistics of the datasets.}
    \label{tab:my_label}
\end{table}

We next describe the datasets that we used.
\begin{enumerate}
\item COLLAB \cite{Morris2020tudataset}. Networks in this dataset are extracted from scientific collaboration networks. 
Each node represents an author of a paper, and two authors are adjacent if they coauthored a paper. 
The networks are sampled from papers in {\em High Energy Physics}, {\em Condensed Matter Physics}, and {\em Astro Physics}, and these form the labels. 

\item IMDB-BINARY and IMDB-MULTI \cite{Morris2020tudataset}.
Each node is an actor, and two nodes are joined if they appear in the same movie. 
The label of a network is the genre of the action movie. 
IMDB-BINARY dataset includes {\em Action} and {\em Romance} movies; IMDB-MULTI dataset includes {\em Comedy}, {\em Romance}, and {\em Science Fiction} movies. 

\item Github Stargazers  \cite{karateclub}. This is a network extracted from GitHub, where each node is a user, and two users are adjacent if they follow the same project. Networks are labeled as belonging to {\em web development} or {\em machine learning} projects.

\item Deezer Ego network \cite{karateclub}. The networks are extracted from European Deezer users, where two users are adjacent if they follow the same artist. Each graph is sampled from users of the same gender; {\em male} and {\em female} are selected in their sampling and form labels.

\item Survivor and Big Brother \cite{CON}. We consider datasets from two social game television shows: \textit{Survivor} and \textit{Big Brother}. This dataset is derived from the episodes of the shows, where the players will vote to remove each other at the end of each episode. The player with the highest number of votes is removed. The co-voting network of each season for the shows forms a directed network, with nodes representing players and directed edges corresponding to votes. We simplify these networks by taking unweighted and undirected edges between players whenever there is a voting between them. Each graph has one of the two labels that indicates the shows, either Survivor or Big Brother. 
\end{enumerate}

To compute the classification accuracy fairly, we use 10-fold cross-validation. For each dataset, we split into 10-folds using the {\em stratified shuffle split} strategy; that is, each partition preserves the percentage of samples of each class. We repeat the experiments ten times and report the mean and the standard deviation of the accuracy of the resulting 100 classifiers. We follow a similar setup of evaluation method to experiment in \cite{Morris2020tudataset}. Primarily, we experiment using $C$-SVM with a linear kernel and $\ell_2$-penalty, and optimize the results from $C\in\{10^{-3}, 10^{-2}, \dots, 10^{3}\}$. We use the term $\ell_2$-penalty to keep consistency as in \cite{sklearn}; some literature uses the term $\ell_i$-regularization, which is the same. For the feature vectors of each graph data, we compare the classification ability between using the clique profile with or without appending the global clustering coefficient of the graphs. 

We report the accuracy score using linear $C$-SVM with $\ell_2$-penalty in Table~\ref{table: main}.  
We use $\mathcal C_k$ and $\mathcal D_k$ to indicate the input feature vector using $\calC_k(G)$ and $\calD_k(G)$, respectively, where $\mathcal{D}_k(G)$ denotes the concatenation of the global clustering coefficient to $\mathcal C_k(G)$.

\begin{table}[ht]
\begin{center}
\begin{tabular}{|c|c|c|c|c|c|c|}\hline
 &   COLLAB  &   \makecell{IMDB-\\BINARY}  &   \makecell{IMDB-\\MULTI}  &   \makecell{Github-\\Stargaers}  &   \makecell{Deezer Ego\\Nets}  &   \makecell{Survivor \&   \\ Big Brother} \\\hline
$\mathrm{acc}$  &   59.98\tiny{$\pm$1.57} &   58.03\tiny{$\pm$4.20} &   38.67\tiny{$\pm$2.88} &   51.48\tiny{$\pm$1.13}&   53.59\tiny{$\pm$1.48}&   74.29\tiny{$\pm$7.69}\\\hline
$\mathrm{cc}$  &   63.63\tiny{$\pm$1.92} &   60.55\tiny{$\pm$4.91} &   39.09\tiny{$\pm$2.77}  &55.58\tiny{$\pm$1.06} &   50.07\tiny{$\pm$1.56} &   73.28\tiny{$\pm$6.48}\\\hline
$\mathcal{C}_{4}$ &   {\bf 69.66\tiny{$\pm$1.76}} &   70.06\tiny{$\pm$4.57} &   44.53\tiny{$\pm$3.01} &   54.37\tiny{$\pm$0.86} &   51.59\tiny{$\pm$1.60} &   {\bf 53.64\tiny{$\pm$11.64}}\\
$\mathcal{C}_{5}$ &   64.90\tiny{$\pm$1.87} &   {\bf 70.37\tiny{$\pm$4.11}} &   46.97\tiny{$\pm$3.61} &   54.67\tiny{$\pm$0.94} &   50.96\tiny{$\pm$1.45} &   52.82\tiny{$\pm$10.52}\\
$\mathcal{C}_{6}$ &   67.35\tiny{$\pm$2.11} &   69.91\tiny{$\pm$4.68} &   {\bf 47.44\tiny{$\pm$3.32}} &   54.66\tiny{$\pm$0.88} &   52.38\tiny{$\pm$1.30} &   52.89\tiny{$\pm$12.31}\\
$\mathcal{C}_{7}$ &   {\bf 69.11\tiny{$\pm$1.69}} &   70.76\tiny{$\pm$4.23} &   46.83\tiny{$\pm$3.45} &   54.64\tiny{$\pm$1.19} &   51.49\tiny{$\pm$1.73} &   50.09\tiny{$\pm$11.75}\\
$\mathcal{C}_{8}$ &   68.05\tiny{$\pm$1.88} &   70.50\tiny{$\pm$4.02} &   48.22\tiny{$\pm$3.50} &   54.67\tiny{$\pm$0.95} &   51.40\tiny{$\pm$1.50} &   {\bf 52.30\tiny{$\pm$12.09}}\\
$\mathcal{C}_{9}$ &   67.74\tiny{$\pm$2.03} &   {\bf 71.10\tiny{$\pm$4.05}} &   48.83\tiny{$\pm$3.26} &   54.72\tiny{$\pm$1.02} &   51.30\tiny{$\pm$1.56} &   51.19\tiny{$\pm$11.92}\\
$\mathcal{C}_{10}$ &   68.13\tiny{$\pm$1.91} &   70.68\tiny{$\pm$4.26} &  {\bf 49.41\tiny{$\pm$3.77}} &   54.77\tiny{$\pm$0.91} &   51.40\tiny{$\pm$1.43} &   51.11\tiny{$\pm$11.60}\\
\hline$\mathcal{D}_{4}$ &   67.23\tiny{$\pm$2.09} &   69.16\tiny{$\pm$4.34} &   45.13\tiny{$\pm$3.85} &   58.23\tiny{$\pm$1.25} &   51.16\tiny{$\pm$1.29} &   86.08\tiny{$\pm$6.68}\\
$\mathcal{D}_{5}$ &   68.97\tiny{$\pm$1.90} &   70.75\tiny{$\pm$4.02} &   46.71\tiny{$\pm$4.31} &   58.44\tiny{$\pm$1.12} &   50.64\tiny{$\pm$1.37} &   83.98\tiny{$\pm$7.80}\\
$\mathcal{D}_{6}$ &   68.82\tiny{$\pm$1.93} &   70.46\tiny{$\pm$4.17} &   47.39\tiny{$\pm$3.81} &   58.42\tiny{$\pm$1.56} &   51.39\tiny{$\pm$1.38} &   83.50\tiny{$\pm$6.76}\\
$\mathcal{D}_{7}$ &   68.60\tiny{$\pm$1.77} &   71.53\tiny{$\pm$4.69} &   47.39\tiny{$\pm$3.40} &   58.51\tiny{$\pm$1.38} &   50.98\tiny{$\pm$1.75} &   83.41\tiny{$\pm$6.79}\\
$\mathcal{D}_{8}$ &   68.08\tiny{$\pm$2.10} &   71.10\tiny{$\pm$4.06} &   48.33\tiny{$\pm$3.79} &   58.46\tiny{$\pm$1.32} &   50.94\tiny{$\pm$1.41} &   83.71\tiny{$\pm$7.00}\\
$\mathcal{D}_{9}$ &   68.26\tiny{$\pm$1.95} &   71.41\tiny{$\pm$3.92} &   49.02\tiny{$\pm$4.06} &   58.44\tiny{$\pm$1.51} &   51.11\tiny{$\pm$1.28} &   84.25\tiny{$\pm$7.33}\\
$\mathcal{D}_{10}$ &   68.91\tiny{$\pm$2.15} &   71.51\tiny{$\pm$4.80} &   49.20\tiny{$\pm$3.51} &   58.49\tiny{$\pm$1.01} &   51.05\tiny{$\pm$1.90} &   84.49\tiny{$\pm$6.60}\\
\hline\cite{Morris2020tudataset} &   N/A &   59.8 \tiny{$\pm$ 1.1} &   39.5\tiny{$\pm$ 0.7} &   N/A&   N/A&   N/A \\\hline \end{tabular}
\captionsetup{margin=1cm}
\end{center}
\caption{The prediction accuracy percentage of linear $C$-SVM with $\ell_2$-penalty. 
Each row with $\calC_k$ indicates the results with $\calC_k(G)$ as an input for each network $G$, and each row with $\calD_k$ indicates $\calD_k(G)$ as the input. 
The last row reports the classification results using graphlet kernel from \cite{Morris2020tudataset}, where N/A stands for not available, as they are not recorded in the paper. Due to the small size of the Survivor \&   Big Brother dataset, the standard deviations of the results tend to be larger. 
The bold numbers indicate examples that make $\mathcal{C}_k$ non-unimodal.}
\label{table: main}
\end{table}

For additional corroboration of our results, see Table~\ref{table: supp}. We also run our experiments using linear $C$-SVM with $\ell_1$-penalty, $C$-SVM with an RBF kernel, and 2-layer multi-layer perceptron neural networks with the number of neurons optimized from $\{\lfloor r\cdot k\rfloor: 4\leq k\leq 10, r\in\{0.7, 1, 1.3\}\}$, where $k$ corresponds to the $k$-clique profile and $r$ stands for a ratio. We use Scikit-Learn 1.3.0 for all classifiers mentioned above; see \cite{sklearn} for details of these algorithms. 

\begin{table}[ht]
\begin{center}
\tiny\begin{tabular}{|c|c|c|c|c|c|c|c|c|c|}\hline
 & \multicolumn{3}{c|}{COLLAB}  & \multicolumn{3}{c|}{\makecell{IMDB-\\BINARY}}  & \multicolumn{3}{c|}{\makecell{IMDB-\\MULTI}} \\\hline
 & \makecell{SVM-$\ell_1$} & \makecell{SVM-RBF} & MLP  & \makecell{SVM-$\ell_1$} & \makecell{SVM-RBF} & MLP  & \makecell{SVM-$\ell_1$} & \makecell{SVM-RBF} & MLP \\\hline
$\mathcal{C}_{4}$  & 67.64\tiny{$\pm$1.8} & 64.85\tiny{$\pm$2.3} & 70.38\tiny{$\pm$2.0} & 70.24\tiny{$\pm$4.9} & 70.47\tiny{$\pm$4.4} & 62.89\tiny{$\pm$7.8} & 44.34\tiny{$\pm$3.0} & 48.11\tiny{$\pm$3.2} & 38.95\tiny{$\pm$2.5}\\
$\mathcal{C}_{5}$  & 63.13\tiny{$\pm$2.0} & 66.45\tiny{$\pm$2.3} & 70.85\tiny{$\pm$2.4} & 70.22\tiny{$\pm$4.6} & 70.40\tiny{$\pm$4.3} & 64.99\tiny{$\pm$5.2} & 45.64\tiny{$\pm$3.1} & 49.79\tiny{$\pm$3.3} & 40.44\tiny{$\pm$6.1}\\
$\mathcal{C}_{6}$  & 63.83\tiny{$\pm$1.8} & 67.00\tiny{$\pm$2.2} & 70.96\tiny{$\pm$2.0} & 69.75\tiny{$\pm$4.2} & 71.45\tiny{$\pm$4.7} & 66.45\tiny{$\pm$6.3} & 46.48\tiny{$\pm$3.2} & 50.28\tiny{$\pm$3.7} & 42.90\tiny{$\pm$6.2}\\
$\mathcal{C}_{7}$  & 67.20\tiny{$\pm$2.0} & 66.90\tiny{$\pm$2.0} & 70.96\tiny{$\pm$1.8} & 70.84\tiny{$\pm$4.0} & 72.47\tiny{$\pm$4.3} & 67.21\tiny{$\pm$4.6} & 46.32\tiny{$\pm$2.6} & 50.19\tiny{$\pm$4.0} & 43.67\tiny{$\pm$5.5}\\
$\mathcal{C}_{8}$  & 67.75\tiny{$\pm$1.9} & 67.03\tiny{$\pm$1.8} & 70.87\tiny{$\pm$1.7} & 70.57\tiny{$\pm$4.0} & 71.76\tiny{$\pm$4.5} & 67.53\tiny{$\pm$4.7} & 46.01\tiny{$\pm$2.8} & 50.44\tiny{$\pm$3.4} & 45.22\tiny{$\pm$5.2}\\
$\mathcal{C}_{9}$  & 67.52\tiny{$\pm$2.1} & 66.74\tiny{$\pm$1.8} & 70.67\tiny{$\pm$2.1} & 70.41\tiny{$\pm$4.4} & 71.60\tiny{$\pm$4.1} & 67.50\tiny{$\pm$4.3} & 47.75\tiny{$\pm$3.1} & 50.18\tiny{$\pm$3.5} & 45.45\tiny{$\pm$5.1}\\
$\mathcal{C}_{10}$  & 67.56\tiny{$\pm$1.9} & 66.70\tiny{$\pm$2.0} & 70.71\tiny{$\pm$1.9} & 70.21\tiny{$\pm$4.3} & 71.36\tiny{$\pm$4.1} & 67.37\tiny{$\pm$4.7} & 47.77\tiny{$\pm$2.9} & 49.96\tiny{$\pm$3.6} & 47.06\tiny{$\pm$5.4}\\
\hline$\mathcal{D}_{4}$  & 65.15\tiny{$\pm$1.9} & 66.85\tiny{$\pm$2.0} & 69.51\tiny{$\pm$3.2} & 69.11\tiny{$\pm$4.0} & 70.57\tiny{$\pm$4.0} & 62.28\tiny{$\pm$8.3} & 45.16\tiny{$\pm$3.3} & 49.15\tiny{$\pm$3.4} & 38.96\tiny{$\pm$4.2}\\
$\mathcal{D}_{5}$  & 65.73\tiny{$\pm$1.8} & 67.29\tiny{$\pm$1.9} & 70.87\tiny{$\pm$2.4} & 69.25\tiny{$\pm$5.1} & 70.61\tiny{$\pm$4.2} & 65.16\tiny{$\pm$6.3} & 46.21\tiny{$\pm$3.6} & 49.67\tiny{$\pm$3.7} & 41.80\tiny{$\pm$5.9}\\
$\mathcal{D}_{6}$  & 67.09\tiny{$\pm$2.1} & 67.57\tiny{$\pm$2.0} & 71.25\tiny{$\pm$2.2} & 70.07\tiny{$\pm$4.4} & 72.16\tiny{$\pm$4.4} & 66.41\tiny{$\pm$5.2} & 47.61\tiny{$\pm$3.3} & 50.00\tiny{$\pm$3.9} & 42.83\tiny{$\pm$5.2}\\
$\mathcal{D}_{7}$  & 68.20\tiny{$\pm$2.2} & 67.74\tiny{$\pm$2.0} & 71.69\tiny{$\pm$1.8} & 71.58\tiny{$\pm$4.6} & 71.67\tiny{$\pm$4.2} & 67.05\tiny{$\pm$5.4} & 47.19\tiny{$\pm$3.7} & 49.89\tiny{$\pm$3.7} & 44.34\tiny{$\pm$5.3}\\
$\mathcal{D}_{8}$  & 67.66\tiny{$\pm$2.0} & 67.78\tiny{$\pm$1.8} & 71.61\tiny{$\pm$2.0} & 71.36\tiny{$\pm$4.1} & 71.80\tiny{$\pm$4.3} & 67.39\tiny{$\pm$4.2} & 46.19\tiny{$\pm$3.9} & 49.37\tiny{$\pm$4.0} & 45.59\tiny{$\pm$5.9}\\
$\mathcal{D}_{9}$  & 67.73\tiny{$\pm$2.1} & 67.94\tiny{$\pm$1.7} & 71.74\tiny{$\pm$2.0} & 70.72\tiny{$\pm$4.4} & 71.82\tiny{$\pm$4.0} & 67.26\tiny{$\pm$4.7} & 48.69\tiny{$\pm$3.7} & 50.72\tiny{$\pm$3.8} & 47.13\tiny{$\pm$6.0}\\
$\mathcal{D}_{10}$  & 67.89\tiny{$\pm$1.9} & 67.65\tiny{$\pm$2.1} & 71.84\tiny{$\pm$2.1} & 70.89\tiny{$\pm$4.6} & 72.10\tiny{$\pm$4.2} & 67.42\tiny{$\pm$4.7} & 48.85\tiny{$\pm$3.7} & 50.32\tiny{$\pm$3.7} & 48.13\tiny{$\pm$5.6}\\
\hline\end{tabular}
\end{center}
\caption{Additional corroboration of results in Table~\ref{table: main}. 
For each dataset, the second row indicates the machine for the classification, where SVM-$\ell_1$, SVM-RBF, and MLP correspond to the classification results from linear $C$-SVM with $\ell_1$-penalty, $C$-SVM with an RBF kernel, and multilayer perceptron neural networks. }
\label{table: supp}
\end{table}

We observe better accuracy in the IMDB-BINARY and IMDB-MULTI datasets than in the benchmark. Though some results show that clique counting cannot classify the labels in the dataset, results of some datasets still verify our hypothesis that clique profiles possess a strong classification ability. For the COLLAB dataset, we did not find a graphlet-based approach to compare as a benchmark following the same experiment routine. 
In \cite{shervashidze2009efficient}, they reported the accuracy on the COLLAB dataset after randomly flipping 10/20/40\% of edges as a simulation of noise with an accuracy of $72.84_{\pm0.28}\%$. 

We observed that increasing $k$ in the $k$-clique profile does not necessarily improve accuracy, yet the accuracy results do not appear unimodal. This situation holds for both $\calC_k$'s and $\calD_k$'s.  We may expect that appending the global clustering coefficient necessarily increases classification accuracy; however, observe that there are a few cases where the accuracy using $\calD_k$ is lower than $\calC_k$.
For instance, $k\in\{4,7\}$ in COLLAB dataset, $k=6$ in Deezer Ego Nets dataset, and some other cases have only a minor difference. 

\section{Discussion and Future Work}

We introduced clique profiles as a fast, elementary measure of graph similarity. We compared clique profiles in various social networks and found them to be accurate separators in many labeled networked datasets. The advantage of using cliques versus full graph profiles or deep learning methods is that they are computationally less expensive.

Applying our approach to more social network datasets would be interesting in future work. While our methods are less applicable to sparse networks with few cliques, one direction would be to consider profiles of sparse subgraphs such as trees to measure graph similarity. Another direction would be to extend clique profiles to hypergraphs, which are useful models for higher-order structures in networks. Implementing Pivoter makes computing the node-wise ($k$-)clustering coefficients more feasible. Another direction is to investigate the effects of clustering coefficients on the node classification problems. 

\section{Acknowledgments}
Research supported by a grant of the first author from NSERC.


\begin{thebibliography}{99}

\bibitem{shford2022online} J.R.\ Ashford, L.D.\ Turner, R.M.\ Whitaker, A.\ Preece, D.\ Felmlee, Understanding the characteristics of COVID-19 misinformation communities through graphlet analysis, \emph{Online Social Networks and Media} \textbf{27} (2022) 100178.

\bibitem{b16} D.R.\ D'Angelo, A.\ Bonato, E.R.\ Elenberg, D.F.\ Gleich, Y.\ Hou, Mining and modeling character networks, In: \emph{Proceedings of Algorithms and Models for the Web Graph}, 2016.

\bibitem{bonato2008course} A. Bonato, \emph{A Course on the Web Graph}, American Mathematical Society, Providence, Rhode Island, 2008.

\bibitem{bonato2022frustum} A.\ Bonato, R.\ Cushman, T.\ Marbach, Z.\ Zhang, An evolving network model from clique extension, In: \emph{Proceedings of The 28th International Computing and Combinatorics Conference}, 2022.

\bibitem{CON} A.\ Bonato, N.\ Eikmeier, D.F.\ Gleich, R.\ Malik, Centrality in dynamic competition networks, In: \emph{Proceedings of Complex Networks}, 2019.

\bibitem{bonato2014dimensionality} A.\ Bonato, D.F.\ Gleich, M.\ Kim, D.\ Mitsche, P.\ Pra\l{}at, A.\ Tian, S.J.\ Young, Dimensionality matching of social networks using motifs and eigenvalues, \emph{PLOS ONE} \textbf{9}(9):e106052, 2014.

\bibitem{borgwardt2020graph} K.\ Borgwardt, E. Ghisu, F. Llinares-L\'opez, L. O'Bray, B. Rieck.\ Graph kernels: State-of-the-art and future challenges, \emph{Foundations and trends in machine learning}, \textbf{13} (2020) 531--712.

\bibitem{feng2022motif} B.\ Feng, Y.\ Yang, L.\ Zhang, S.\ Xue, X.\ Xie, J.\ Wang, G.\ Xie, Motif importance measurement based on multi-attribute decision, \emph{Journal of Complex Networks} \textbf{10} (2022) cnac023.

\bibitem{fox2020finding} J.\ Fox, T.\ Roughgarden, C.\ Seshadhri, F.\ Wei, N.\ Wein, Finding cliques in social networks: A new distribution-free model, \emph{SIAM journal on computing} \textbf{49} (2020) 448--464.

\bibitem{jain2020power} S.\ Jain, C.\ Seshadhri, The power of pivoting for exact clique counting, In: \emph{Proceedings of the 13th International Conference on Web Search and Data Mining}, 2020.

\bibitem{hurshman2012model} J.\ Janssen, M.\ Hurshman, N.\ Kalyaniwalla, Model selection for social networks using graphlets, \emph{Internet
Mathematics} \textbf{8} (2012) 338--363.

\bibitem{lawford2020cliques} S.\ Lawford, Y.\ Mehmeti, Cliques and a new measure of clustering: With application to US domestic airlines, \emph{Physica A: Statistical Mechanics and its Applications} \textbf{560} (2020) 125158.

\bibitem{leskovec2005graphs} J.\ Leskovec, J.\ Kleinberg, C.\ Faloutsos, Graphs over time: densification laws, shrinking diameters and possible explanations, In: \emph{Proceedings of the eleventh ACM SIGKDD international conference on Knowledge discovery in data mining}, 2005.

\bibitem{snap} J.\ Leskovec, A.\ Krevl, {SNAP Datasets}: {Stanford} Large Network Dataset Collection, {\url{http://snap.stanford.edu/data}}, 2014.

\bibitem{milo2002network} R.\ Milo, Ron S.\ Shen-Orr, S.\ Itzkovitz, N.\ Kashtan, D.\ Chklovskii, U.\ Alon, Network motifs: simple building blocks of complex networks, \emph{Science} \textbf{298} (2002) 824--827.

\bibitem{Morris2020tudataset} C.\ Morris, N.M.\ Kriege, F.\ Bause, K.\ Kersting, P.\ Mutzel, M.\ Neumann, TUDataset: A collection of benchmark datasets for learning with graphs, In: \emph{Proceedings of ICML 2020 Workshop on Graph Representation Learning and Beyond}, 2020.

\bibitem{sklearn} F.\ Pedregosa, G.\ Varoquaux, A.\ Gramfort, V.\ Michel, B.\ Thirion, O.\ Grisel, M.\ Blondel, P.\ Prettenhofer, R.\ Weiss, V.\ Dubourg, J.\ Vanderplas, A.\ Passos, D.\ Cournapeau, M.\ Brucher, M.\ Perrot, E.\ Duchesnay, Scikit-learn: machine learning in Python,  \emph{{Journal of Machine Learning Research}} \textbf{12} (2011) 2825--2830. 

\bibitem{Pi2023cliquedensification} H.\ Pi, K.\ Burghardt, A.G.\ Percus, K.\ Lerman, Clique densification in networks, \emph{Phys. Rev. E} \textbf{107} (2023) L042301.

\bibitem{prvzulj2007biological} N.\ Pr{\v{z}}ulj, Biological network comparison using graphlet degree distribution, \emph{Bioinformatics} (2007) \textbf{23} e177--e183.

\bibitem{ribeiro2021survey} P.\ Ribeiro, P.\ Paredes, M.\ Silva, D.\ Aparicio and F.\ Silva, A survey on subgraph counting: concepts, algorithms, and applications to network motifs and graphlets, \emph{ACM Computing Surveys (CSUR)} \textbf{54} (2021) 1--36.  

\bibitem{karateclub} B.\ Rozemberczki, O.\ Kiss, R.\ Sarkar, Karate Club: An API oriented open-source Python framework for unsupervised learning on graphs, In: \emph{Proceedings of the 29th ACM International Conference on Information and Knowledge Management}, 2020.

\bibitem{shervashidze2009efficient} N.\ Shervashidze, S.V.N. Vishwanathan, T.\ Petri, K.\ Mehlhorn, K.\ Borgwardt, Efficient graphlet kernels for large graph comparison, \emph{Artificial intelligence and statistics} (2009) 488--495.

\bibitem{sinha2022impact} S.\ Sinha, S.\ Bhattacharya, S.\ Roy, Impact of second-order network motif on online social networks, \emph{The Journal of Supercomputing} \textbf{78} (2022) 5450--5478.

\bibitem{watts1998collective} D.J.\ Watts, S.H.\ Strogatz, Collective dynamics of `small-world' networks, \emph{Nature} \textbf{393} (1998) 440--442.

\bibitem{west} D.B.\ West, \emph{Introduction to Graph Theory, 2nd edition}, Prentice Hall, 2001.

\bibitem{yanardag2015deep} P.\ Yanardag, SVN Vishwanathan, Deep graph kernels, In: \emph{Proceedings of the 21st ACM SIGKDD international conference on knowledge discovery and data mining}, 2015.

\bibitem{yin2018higher} H.\ Yin, A.R.\ Benson, J.\ Leskovec, Higher-order clustering in networks, \emph{Physical Review E} \textbf{97} (2018) 052306.

\bibitem{zhao2023intrinsic} H.\ Zhao, C.\ Shao,Z.\ Shi, S.\ He, Shibo, Z.\ Gong, The intrinsic similarity of topological structure in biological neural networks, \emph{IEEE/ACM Transactions on Computational Biology and Bioinformatics} (2023).

\end{thebibliography}
\end{document}